# Numerical simulations of anomalous diffusion

**Mariusz Ciesielski and Jacek Leszczynski**
*Institute of Mathematics and Computer Science, Technical University of Czestochowa*
*Dabrowskiego 73, 42-200 Czestochowa, POLAND*
*e-mail: cmariusz@k2.pcz.czest.pl*

Abstract

In this paper we present numerical methods – finite differences and finite elements – for solution of partial differential equation of fractional order in time for one-dimensional space. This equation describes anomalous diffusion which is a phenomenon connected with the interactions within the complex and non-homogeneous background. In order to consider physical initial-value conditions we use fractional derivative in the Caputo sense. In numerical analysis the boundary conditions of first kind are accounted and in the final part of this paper the result of simulations are presented.

## 1. Introduction

Anomalous diffusion is a phenomenon strongly connected with the interactions within the complex and non-homogeneous background. This phenomenon is observed in chaotic heat baths [14], diffusion through porous materials [10, 17], amorphous semiconductors [15], nuclear magnetic resonance diffusometry in disordered materials [17], behaviour of polymers in a glass transition [2, 23], particle dynamics inside polymer network [9], and also in eddy flows [25]. Many investigators proposed models which based on the linear and non-linear forms of differential equations. Such models can simulate anomalous diffusion but they do not reflect its real behaviour. The authors in Refs. [1, 3, 16, 19, 21, 22, 24] use fractional calculus [1, 3, 5, 6, 8] for modelling this type of diffusion. This means that spatio-temporal derivatives in the classical diffusion equation are exchanged by fractional ones. The mathematics of fractional calculus is a natural extension towards integer-order calculus.

In comparison to derivatives of integer order, which depend on the local behaviour of the function, derivatives of fractional order accumulate the whole history of this function and it is so-called the memory effect. The memory effect leads to many applications of differential equations of fractional order for both ordinary and partial equations.

## 2. Mathematical model

In this paper we consider the fractional partial differential equation in the following form

$$\frac{{}^c\partial^a}{\partial t^a}u(x,t) = k_a \frac{\partial^2}{\partial x^2}u(x,t),\ t \geq 0,\ x \in \mathbf{R}, \quad (1)$$

where $u(x,t)$ is a field variable (e.g. the probability density of diffusive displacements $x$ in a time $t$), ${}^C\partial^a/\partial t^a$ is the fractional derivative in the Caputo sense, $a$ is the real order of this operator, $k_a$ is a coefficient of generalised (anomalous) diffusion $[m^2/s^a]$. With regard to Eqn (1) we obtain the classical diffusion equation for $a = 1$, e.g. the heat transfer equation. If $a = 2$, on the other hand, the wave equation may note. Therefore we assume variations of the parameter $a$ in the range $0 < a \leq 2$. Analysing behaviour of the parameter $a$ in Eqn (1) we notice a relaxation process (*sub-diffusion*) when $0 < a < 1$. If $1 < a < 2$ a relaxation - oscillation process (*super-diffusion*) may occur [1, 19, 20].

Fractional calculus involves different definitions of the fractional operator as the Riemann-Liouville fractional derivative, the Caputo derivative, the Grünwald-Letnikov derivative, the Riesz derivative and also the Weyl-Marchaud derivative [5, 6, 8].

We introduce the definition of operator ${}^C_0D^a_t$ as the left-sided Caputo derivative and for $a \notin \mathbf{N}$ we have

$${}^C_0D^a_t\boldsymbol{j}(t,\cdot) = \frac{{}^c\partial^a}{\partial t^a}\boldsymbol{j}(t,\cdot) := \frac{1}{\Gamma(m+1-a)}\int_0^t \frac{\frac{\partial^{m+1}}{\partial x^{m+1}}\boldsymbol{j}(x,\cdot)}{(t-x)^{a-m}}dx, \quad (2)$$

where $m = [a]$, $[\cdot]$ denotes an integer part of a real number. However, for $a \in \mathbf{N}$ uses the classical definitions of integer derivatives. Note that the operator ${}_0D^a_t$ is defined in the Riemann-Liouville sense as

$${}_0D^a_t\boldsymbol{j}(t,\cdot) = \frac{\partial^a}{\partial t^a}\boldsymbol{j}(t,\cdot) := \frac{1}{\Gamma(m+1-a)}\frac{\partial^{m+1}}{\partial t^{m+1}}\int_0^t \frac{\boldsymbol{j}(x,\cdot)}{(t-x)^{a-m}}dx. \quad (3)$$

Mathematical modelling based on anomalous diffusion leads to the differential equations of fractional order and to necessity of formulation of initial conditions to these equations. For the standard diffusion equation ($a = 1$), which is the equation in the parabolic form, one initial function is sufficient, whereas for the standard wave equation ($a = 2$), which is the equation in the hyperbolic form, the time derivative of function is added. For the intermediate scheme considered here, i.e. ($1 < a < 2$), we need two initial functions as in the case of $a = 2$. Physical systems involving fractional derivatives allow us to utilize a special form of initial-value conditions, which contain $\boldsymbol{j}(0^+,\cdot)$, $(\partial/\partial t)\boldsymbol{j}(0^+,\cdot)$ etc. Unfortunately, the fractional differential equations with the Riemann-Liouville operator lead to initial conditions containing the limit values at $t = 0$, i.e. $(\partial^{a-1}/\partial t^{a-1})\boldsymbol{j}(0^+,\cdot)$, $(\partial^{a-2}/\partial t^{a-2})\boldsymbol{j}(0^+,\cdot)$ etc. The initial-value problem with such initial conditions is solved mathematically, but these solutions are useless practically (for such a type of initial conditions the physical interpretations is unknown). Caputo in [11] proposed the definition (2) which enables the initial conditions as the same form as for differential equations of integer order. According to fractional calculus [1, 3, 6, 13] between the Caputo and the Riemann-Liouville derivatives is following form

$$^C_0D^a_t \boldsymbol{j}(\boldsymbol{t},\cdot) = {_0D^a_t}\boldsymbol{j}(\boldsymbol{t},\cdot) - \sum_{k=0}^{m} \frac{\boldsymbol{t}^{k-\boldsymbol{a}}}{\Gamma(k-\boldsymbol{a}+1)} \frac{\partial^k}{\partial \boldsymbol{t}^k}\boldsymbol{j}(0^+,\cdot). \quad (4)$$

Additional interesting properties of the fractional operators (e.g. the composition and chain rules) one can find in literature [5, 6, 8, 13, 18].

In this paper we will considered Eqn (1) in 1D domain $\Omega$: $0 \leq x \leq L$ with boundary-value conditions of first kind (the Dirichlet conditions) as

$$\begin{cases} x=0: & u(0,t)=g_0(t) \\ x=L: & u(L,t)=g_L(t) \end{cases} \quad t>0 \quad (5)$$

and with initial-value conditions

$$\begin{cases} u(x,t)\big|_{t=0} = p_0(x) \\ \dfrac{\partial}{\partial t}u(x,t)\bigg|_{t=0} = p_1(x), \text{ for } 1<\boldsymbol{a}\leq 2 \end{cases}. \quad (6)$$

Boundary conditions of other kinds (i.e. the von Neumann and the Robin conditions) allow us to treat the fractional partial equation in similar way as for the classical equations involving the second spatial derivative.

Analytical solution of Eqn (1), which does not contain boundary conditions, uses Green or Fox functions [1, 6, 16, 19]. Nevertheless, we need to solve numerically Eqn (1) when e.g. additional non-linear term occurs. In this general case we adopt boundary conditions, which can vary over time. Different numerical methods, with their advantages and disadvantages, used for solution of fractional differential equations one can find in Refs. [5, 6, 12, 13, 18].

## 3. Numerical algorithms

In this section we concentrate on description of numerical methods used for solution of Eqn (1). In a boundary-initial problem one defines two kinds of grids: spatial and temporal.

We divide the spatial domain $\Omega = [0, L]$ into the uniform mesh with $N + 1$ nodes $x_i = ih$ for $i = 0,\ldots,N$, where $h = L/N$. The time-interval $[0, T]$ is divided into $F$ subintervals where the subinterval length equals $\Delta t = T/F$ and the time-nodes are $t_f = f \times \Delta t$ for $f = 0,\ldots,F$.

In case where $\boldsymbol{a}$ is an integer number, one can find in [4, 7] the finite difference scheme for the diffusion equation ($\boldsymbol{a} = 1$) and the wave equation ($\boldsymbol{a} = 2$).

### 3.1. Discrete forms of fractional derivatives

In this subsection we introduce another definition of the fractional derivative. With regard to [5, 6, 8] we define the fractional derivative if the Grünwald-Letnikov sense

$$^{GL}_0D^a_t \boldsymbol{j}(\boldsymbol{t},\cdot) := \lim_{\Delta t \to 0} (\Delta t)^{-\boldsymbol{a}} \sum_{j=0}^{[t/\Delta t]} (-1)^j \binom{\boldsymbol{a}}{j} \boldsymbol{j}(\boldsymbol{t}-j\Delta t,\cdot). \quad (7)$$

The definition of operator in the Grünwald-Letnikov sense (7) is equivalent to the definition of operator in the Riemann-Liouville sense (3) [5, 6]. Nevertheless the Grünwald-Letnikov operator is more flexible and most straightforward in numerical calculations. The Grünwald-Letnikov operator (7) is approximated within the interval $[0, \boldsymbol{t}]$ with the subinterval length $\Delta t$ as

$$^{GL}_0D^a_t \boldsymbol{j}(\boldsymbol{t},\cdot) \approx \sum_{j=0}^{[t/\Delta t]} c_j^{(a)} \boldsymbol{j}(\boldsymbol{t}-j\Delta t,\cdot) \quad (8)$$

where $c_j^{(a)}$ are Grünwald-Letnikov coefficients defined as

$$c_j^{(a)} = (\Delta t)^{-\boldsymbol{a}}(-1)^j \binom{\boldsymbol{a}}{j}, \text{ for } j=0,1,\ldots . \quad (9)$$

Using the recurrence relationship [6]

$$c_0^{(a)} = (\Delta t)^{-\boldsymbol{a}}, \quad c_j^{(a)} = \left(1 - \frac{1+\boldsymbol{a}}{j}\right) c_{j-1}^{(a)}, \text{ for } j=1,2,\ldots \quad (10)$$

we can compute the coefficients in a simple way. For $j = 1$ we have $c_1^{(a)} = -\boldsymbol{a}(\Delta t)^{-\boldsymbol{a}}$.

This scheme is simple for computational performance. In Refs. [5, 6, 13] are presented and discussed other numerical schemes for fractional derivatives which also are used for calculation of the coefficients $c_j^{(a)}$. It should be noted that the fractional derivative is represented by the weighted sum over the history. In the upper limit $\boldsymbol{t}$, where the derivative is calculated, it notices significant values of the coefficients $c_j^{(a)}$. If our calculations tend backward to the lower limit 0, smaller values of the coefficients $c_j^{(a)}$ may observed.

### 3.2. Finite difference method (FDM)

Deriving the difference scheme for Eqn (1) we start with a description, which contains initial-boundary conditions in the numerical algorithm. We assume $m = [\boldsymbol{a}]$. Introducing initial conditions (6) we can directly determine values of the function $u$ at first time step $t = t_f$ for $f = 0,\ldots,m$ for every nodes $x_i$ and for $i = 1,\ldots, N-1$

$$\begin{aligned} u(x_i,t_0) &= p_0(ih) \\ u(x_i,t_1) &= p_0(ih) + \Delta t \cdot p_1(ih), \text{ for } m=1 \end{aligned}. \quad (11)$$

Boundary conditions of the first kind (5) are used directly to the boundary nodal values (i.e. to the first and to the last node) $x_0$ and $x_N$ at every moments of time $t = t_f$ for $f = 0,\ldots,F$

$$\begin{aligned} u(x_0,t_f) &= g_0(f \cdot \Delta t) \\ u(x_N,t_f) &= g_L(f \cdot \Delta t) \end{aligned}. \quad (12)$$

Using classical expression for the form of the second derivative at the space occurring in the right-side of Eqn (1) we obtain

$$\frac{\partial^2}{\partial x^2}u(x,t)\bigg|_{x=x_i} = \frac{u(x_{i+1},t)-2u(x_i,t)+u(x_{i-1},t)}{h^2}. \quad (13)$$

Whereas the left-side of Eqn (1) at every moments of time $t = t_f$ for $f = m+1,\ldots,F$ replaces by the fractional differential scheme (8) which includes initial conditions (6). Then we have

$$\left.\frac{^c\partial^a}{\partial t^a}u(x,t)\right|_{\substack{t=t_f \\ x=x_i}}$$
$$=\sum_{j=0}^{f}c_j^{(a)}u(x_i,t_{f-j})-\sum_{k=0}^{m}\frac{(t_f)^{k-a}}{\Gamma(k-a+1)}p_k(x_i) \qquad (14)$$

Let us write Eqn (14) in an „explicit" scheme assuming $t = t_{f-1}$. Substituting (13) and (14) into Eqn (1) we obtain

$$\sum_{j=0}^{f}c_j^{(a)}u(x_i,t_{f-j})-\sum_{k=0}^{m}\frac{(t_f)^{k-a}}{\Gamma(k-a+1)}p_k(x_i)$$
$$=k_a\frac{u(x_{i+1},t_{f-1})-2u(x_i,t_{f-1})+u(x_{i-1},t_{f-1})}{h^2} \qquad (15)$$

Denoting $u_i^j = u(x_i,t_j)$, $p_{k,i} = p_k(x_i)$ we write Eqn (15) in the form

$$\sum_{j=0}^{f}c_j^{(a)}u_i^{f-j}-\sum_{k=0}^{m}\frac{(t_f)^{k-a}}{\Gamma(k-a+1)}p_{k,i}=\frac{k_a}{h^2}\left(u_{i+1}^{f-1}-2u_i^{f-1}+u_{i-1}^{f-1}\right). \qquad (16)$$

Disintegrating partially into terms the first sum in Eqn (16) we obtain

$$(\Delta t)^{-a}\left(u_i^f-au_i^{f-1}\right)+\sum_{j=2}^{f}c_j^{(a)}u_i^{f-j}-\sum_{k=0}^{m}\frac{(t_f)^{k-a}}{\Gamma(k-a+1)}p_{k,i}$$
$$=\frac{k_a}{h^2}\left(u_{i+1}^{f-1}-2u_i^{f-1}+u_{i-1}^{f-1}\right) \qquad (17)$$

and finally we have

$$u_i^f=(\Delta t)^a\left[\left(\frac{a}{(\Delta t)^a}-2\frac{k_a}{h^2}\right)u_i^{f-1}+\frac{k_a}{h^2}\left(u_{i+1}^{f-1}+u_{i-1}^{f-1}\right)\right.$$
$$\left.-\sum_{j=2}^{f}c_j^{(a)}u_i^{f-j}+\sum_{k=0}^{m}\frac{(t_f)^{k-a}}{\Gamma(k-a+1)}p_{k,i}\right] \qquad (18)$$

The coefficient occurring at $u_i^{f-1}$ must be positive in order to assure the stability of the „explicit" scheme (18) and therefore we obtain

$$(\Delta t)^a\left(\frac{a}{(\Delta t)^a}-2\frac{k_a}{h^2}\right)\geq 0, \qquad (19)$$

hence yields

$$\Delta t \leq \left(\frac{h^2 a}{2k_a}\right)^{\frac{1}{a}}. \qquad (20)$$

*3.3. Finite element method (FEM)*

Here we present a numerical algorithm on the base on the finite element method. This method consists in discretisation of the considered Eqn (1) over space and over time, which is similar to FDM. Semi-discretisation (over space) of Eqn (1) through Galerkin method [4] reads

$$\mathbf{Ku}+\mathbf{Mu}_t^{(a)}=\mathbf{b}, \qquad (21)$$

where $\mathbf{K}$ is the stiffness matrix, $\mathbf{M}$ is the mass matrix, $\mathbf{u}$ is the vector of field variable $u$, $\mathbf{u}_t^{(a)}$ is the vector of derivatives of fractional order $a$ in time $t$ and $\mathbf{b}$ is the right-hand side vector, whose values are calculated on the basis of boundary conditions. Dimension of the system (21) is equals $N + 1$. We use one-dimensional linear elements and therefore the coefficients of matrixes $\mathbf{K}$ and $\mathbf{M}$ are determined in [4]. Thus we can see that the FEM discretisation in space for Eqn (1) results in a set of ordinary differential equations of fractional order over time.

The vector $\mathbf{u}_t^{(a)}$ at the moment of time $t_f$ can be written as

$$\mathbf{u}_t^{(a)}=\sum_{j=0}^{f}c_j^{(a)}\mathbf{u}^{f-j}-\sum_{k=0}^{m}\frac{(t_f)^{k-a}}{\Gamma(k-a+1)}\mathbf{p}_k$$
$$=(\Delta t)^{-a}\mathbf{u}^f+\sum_{j=1}^{f}c_j^{(a)}\mathbf{u}^{f-j}-\sum_{k=0}^{m}\frac{(t_f)^{k-a}}{\Gamma(k-a+1)}\mathbf{p}_k \qquad (22)$$
$$=(\Delta t)^{-a}\left[\mathbf{u}^f+(\Delta t)^a\sum_{j=1}^{f}c_j^{(a)}\mathbf{u}^{f-j}-(\Delta t)^a\sum_{k=0}^{m}\frac{(t_f)^{k-a}}{\Gamma(k-a+1)}\mathbf{p}_k\right]$$

where $\mathbf{p}_k$ are vectors of initial conditions in every nodes for $k = 0,\ldots,m$. Here we introduce

$$\mathbf{h}^f=-(\Delta t)^a\sum_{j=1}^{f}c_j^{(a)}\mathbf{u}^{f-j} \qquad (23)$$

$$\mathbf{c}^f=(\Delta t)^a\sum_{k=0}^{m}\frac{(t_f)^{k-a}}{\Gamma(k-a+1)}\mathbf{p}_k. \qquad (24)$$

The vector $\mathbf{h}^f$ represents the vector of history and the vector $\mathbf{c}^f$ indicates initial conditions at the moment of time $t_f$, respectively. Substituting (22) into (21) and employing the vectors $\mathbf{u}$ and $\mathbf{b}$ at the moment of time $t_f$, the Eqn (21) takes the following form

$$\mathbf{Ku}^f+(\Delta t)^{-a}\mathbf{M}\left(\mathbf{u}^f-\mathbf{h}^f-\mathbf{c}^f\right)=\mathbf{b}^f. \qquad (25)$$

After mathematical calculations we obtain

$$\left(\mathbf{K}+(\Delta t)^{-a}\mathbf{M}\right)\mathbf{u}^f=(\Delta t)^{-a}\mathbf{M}\left(\mathbf{h}^f+\mathbf{c}^f\right)+\mathbf{b}^f \qquad (26)$$

or in form

$$\left(\mathbf{M}+(\Delta t)^a\mathbf{K}\right)\mathbf{u}^f=\mathbf{M}\left(\mathbf{h}^f+\mathbf{c}^f\right)+(\Delta t)^a\mathbf{b}^f. \qquad (27)$$

According to above expressions we can calculate values $\mathbf{u}^f$.

Using the Dirichlet boundary conditions we modify the first and last rows in the global system of equations (27).

**4. Example of calculations**

On the base of numerical algorithms presented in previous section several examples are computed. Figure 1 and Fig. 2 present the calculations for different values of parameter $a$, for $a \in \langle 0.75, 1.25, 1.5, 1.75\rangle$. In Fig. 1 we can see a set of plots that present the case of solution of Eqn (1) over space at different moments of time. Figure 2 shows solution of Eqn (1) over time in the point $x = 0.5\ m$ of domain $\Omega$. We can observe the relaxation features of Eqn (1) when $a < 1$. For $1 < a < 2$ the relaxation-oscillation features can be noted. In this calculation assumed $L = 1$, $k_a = 1$, initial conditions $p_0(x) = 0$, $p_1(x) = 0$ being constant values in the space and constant boundary conditions $g_0(t) = 40$, $g_L(t) = 20$ in time.

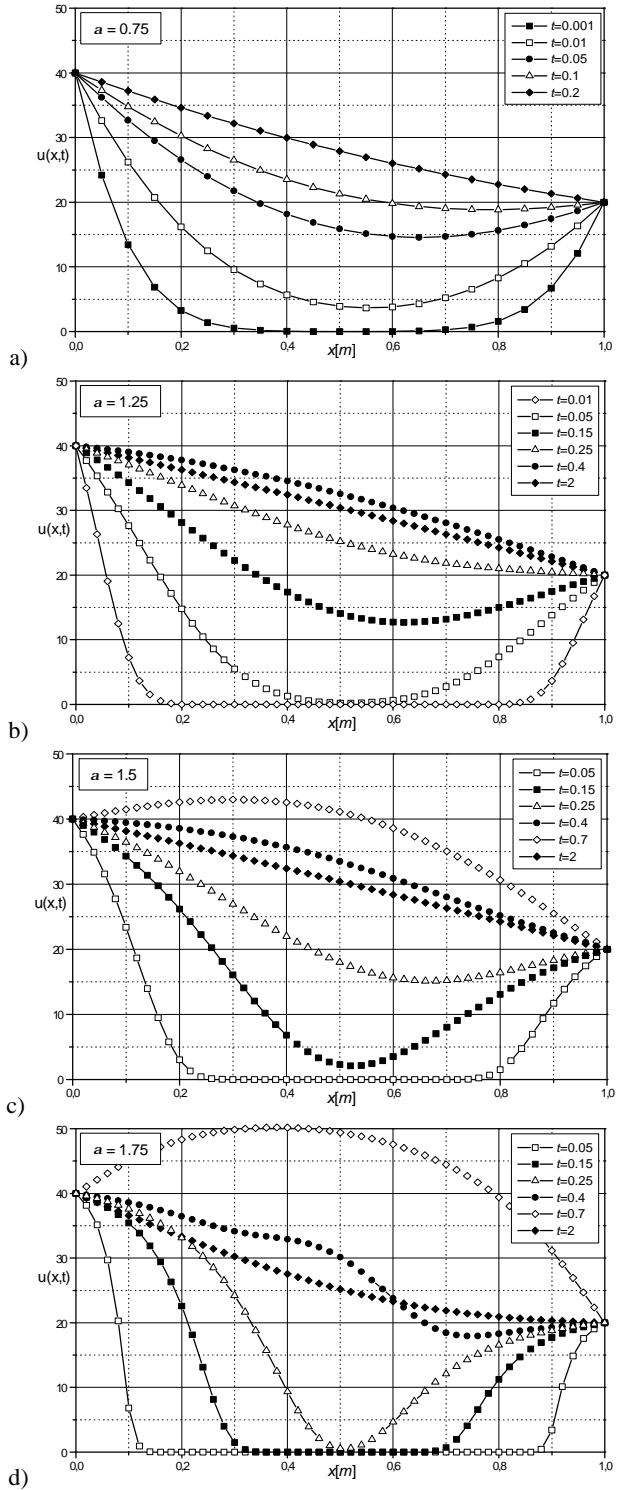

Figure 1: Solution of fractional equation over space for different moments of time and for $a$ = 0.75, 1.25, 1.5 and 1.75.

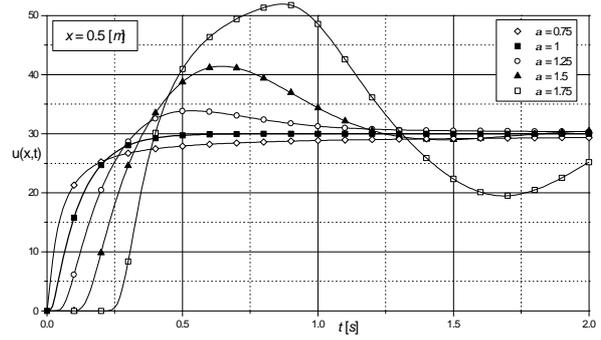

Figure 2: Solution of fractional equation over time for x = 0.5 $m$ and for different values of the parameter $a$.

## 5. Conclusions

We introduced numerical schemes for both FDM and FEM which we successfully used that to perform simulations of anomalous diffusion. We note that the solution of Eqn (1) tends to steady regime for $a \neq 2$.

The fractional derivative has fading memory. A large number of historical data points were made available to each calculation. In schemes FDM (18) and FEM (27) for Eqn (1) we need $f$ - 1 previous values of the function $u$ at time steps that to calculate values of the function $u$ at time $t_f$. This occupies a lot of computational time.

In future researches we will focus on the estimation of the coefficient of anomalous diffusion $k_a$, which depends on physical properties of the system. We will also deal with the numerical problem how to reduce computational time.